# Long-term Pulses of Dynamic Coupling between Solar Hemispheres

D.M. Volobuev, N.G. Makarenko

*Pulkovo Astronomical Observatory, Russian Academy of Sciences, Pulkovskoe sh. 65, St. Petersburg, 196140 Russian Federation* dmitry.volobuev@mail.ru

**Abstract** North-south (N-S) asymmetry of solar activity is a known statistical phenomenon but its significance is difficult to prove or theoretically explain. Here we consider each solar hemisphere as a separate dynamical system connected with the other hemisphere via an unknown coupling parameter. We use a non-linear dynamics approach to calculate the scale-dependent conditional dispersion (CD) of sunspots between hemispheres. Using daily Greenwich sunspot areas, we calculated the Neumann and Pearson chi-squared distances between CDs as indices showing the direction of coupling. We introduce an additional index of synchronization which shows the strength of coupling and allows us to discriminate between complete synchronization and independency of hemispheres. All indices are evaluated in a four-year moving window showing the evolution of coupling between hemispheres. We find that the driver-response interrelation changes between hemispheres have a few pulses during 130 years of Greenwich data with an at least 40 years-long period of unidirectional coupling. These sharp nearly simultaneous pulses of all causality indices are found at the decay of some 11-year cycles. The pulse rate of this new phenomenon of dynamic coupling is irregular: although the first two pulses repeat after 22-year Hale cycles, the last two pulses repeat after three and four 11-year cycles respectively. The last pulse occurs at the decay phase of Cycle 23 so the next pulse will likely appear during the decay of future Cycle 25 or later. This new phenomenon of dynamic coupling reveals additional constraints for understanding and modeling of the long-term solar activity cycles.

**Keywords** Solar Cycle, Sunspots, Statistics

## 1. Introduction

It is now believed that the sunspots are manifestation of the ongoing dynamo inside the solar convection zone. If we carefully examine the dynamo equations and want to reproduce the irregular cycle, we need to stochastically perturb the dynamo equations or need to deal with a nonlinear deterministic dynamo model derived as a "toy" model or as appropriate reduction of the mean field dynamo equations. All these cases have difficulties with reproducing the long-term memory manifested in direct solar indices and proxies as the Gleissberg and Suess cycles (Bushby and Tobias, 2007; Volobuev, 2006). The relatively short length of directly observed indices and the uncertainties in their proxies make it difficult to predict even one decadal sunspot number ahead (Volobuev and Makarenko, 2008) so both the level of determinism and the physical nature of cycles longer than the Hale magnetic cycle remain questionable. Modelling these cycles and the associated grand minima requires making assumptions about the nature of the N-S asymmetry phenomenon (Olemskoy and Kitchatinov, 2013). The possibilities as to why the known mean-field dynamo equations are not able to describe long-term cycles in a unique way may include various parameter ranges which are used in those models, and a number of uncertainties and assumptions about plasma behavior below the photosphere. Different techniques of time series analysis applied to sunspot area data are important as far as these techniques can put additional data-driven constraints on modeling efforts.

N-S asymmetry is a measure of observable difference between solar activity in the northern and southern hemispheres. The oldest asymmetry index is calculated from Greenwich sunspot areas as $(A_N-A_S)/(A_N+A_S)$, here $A_N$ and $A_S$ are the northern and southern hemisphere sunspot areas respectively. The statistical significance of asymmetry is impossible to prove (Carbonell *et al*., 2007) but it is typically considered as a physical phenomenon rather than statistical artifact. Mandal *et al.,* (2017) carried out an independent study of asymmetry using the Kodaikanal data and latitude distributions of sunspots for a particular cycle. They found that the long-term behavior for the height and center of the fitted Gaussian functions revealed some persistence so that asymmetry likely has a long-term cycle similar to the Gleissberg cycle. Such studies prove that this asymmetry is more of a physical phenomena rather a statistical effect. Historical observations during the Maunder minimum manifested that during more than forty years all sunspots were generated in one hemisphere while the other hemisphere was quiet (see *e.g.* Sokoloff and Nesme-Ribes, 1994). The fluctuation of the polar field is believed to be one of the reasons for the N-S asymmetry (*e.g.* Goel and Choudhuri, 2009) while another reason can be different meridional profiles in different hemispheres which may give this asymmetry (Shetye *et al.*, 2015). Recent studies (McIntosh *et al.,* 2015; Javaraiah, 2015) consider the solar dynamo in interacting latitude bands with cross-equator interaction also varying in time. Here we assume the hypothesis (Volobuev and Makarenko, 2016) that the dynamo can act separately in two hemispheres. Considering the Babckock-Leighton dynamo mechanism, there is a problem to transport the poloidal field

Long-term Pulses of Dynamic Coupling between Hemispheres

from polar regions and couple hemispheres across the equator which leads to the so-called solar parity issue (*e.g.* Hotta and Yokoyama, 2010). It is believed that polar field is transported near the tachocline with the meridional cell and there is a problem of coupling these meridional cells between hemispheres. This problem is typically solved by introducing the requirement that the poloidal field should diffuse efficiently so that it is coupled across the equator (Chatterjee, Nandy and Choudhuri, 2004) but parameters remain uncertain so different solutions are possible (Chatterjee and Choudhuri, 2006). If the diffusion rate is high enough the dynamo will not switch to a quadrupole mode but if it is not infinite, there always will be the situation that one hemisphere is the driver and the other is the response system. In this paradigm, the diffusion of the poloidal field is responsible for the coupling. Modelling of this effect is out of the scope for the present work. It would require separate consideration of the poloidal field flow for each hemisphere. The goal of the present work is to evaluate how this coupling behaves with time from the observed sunspot area data using an advanced timeseries processing method. We introduce new indices of dynamical coupling and estimate how the force and direction of coupling varies with time using the approach of non-linear dynamics without any assumptions about the exact mechanism of the dynamo.

The work is organized as follows: Section 2 describes the data and preprocessing algorithm and Section 3 describes the causality detection method. Section 4 illustrates the method with the model example of coupled Van der Pol oscillators; Section 5 shows the implementation of this method to hemispheric sunspot areas. Discussion and conclusions are presented in Section 6.

## 2. Data Description and Preprocessing

A catalog of Greenwich sunspot areas is available at
http://solarscience.msfc.nasa.gov/greenwch.shtml with daily sunspot areas given for the northern and southern hemisphere separately (Figure 1). The normalized asymmetry is calculated as

$$A = <A_N - A_S>/<A_N + A_S> \quad . \quad (1)$$

Here $A_N$ and $A_S$ are the areas of sunspots observed in the northern and southern hemispheres respectively (see *e.g.* Carbonell, Oliver and Ballester, 1993). The power of the normalized asymmetry $A^2$ (Figure 1) is calculated after taking the average of the respective daily sunspot area indices in a 4-year running window. This average allows us to avoid uncertainties with a zero denominator in Equation (1) appearing in spotless days which always appear during the minimum of the 11-year cycle. $A^2$ has a pronounced 11-year cycle which is not in phase with the activity cycle. The absolute asymmetry index $A_N - A_S$ (Temmer *et al.*, 2006) also has an 11-year cycle (Figure 1) without obvious longer cycles. Wavelet spectrum analysis may reveal other multiple cycles in asymmetry including 22-year, 33-year cycles *etc.* (Nagovitisyn and Kuleshova, 2015) but the significance of these cycles is questionable and difficult to prove because of the short period of observations involved which leads to poor statistics. In calculating the causality indices we also use a 4-year running window for processing these indices from daily sunspot areas, which requires an appropriate preliminary smoothing.

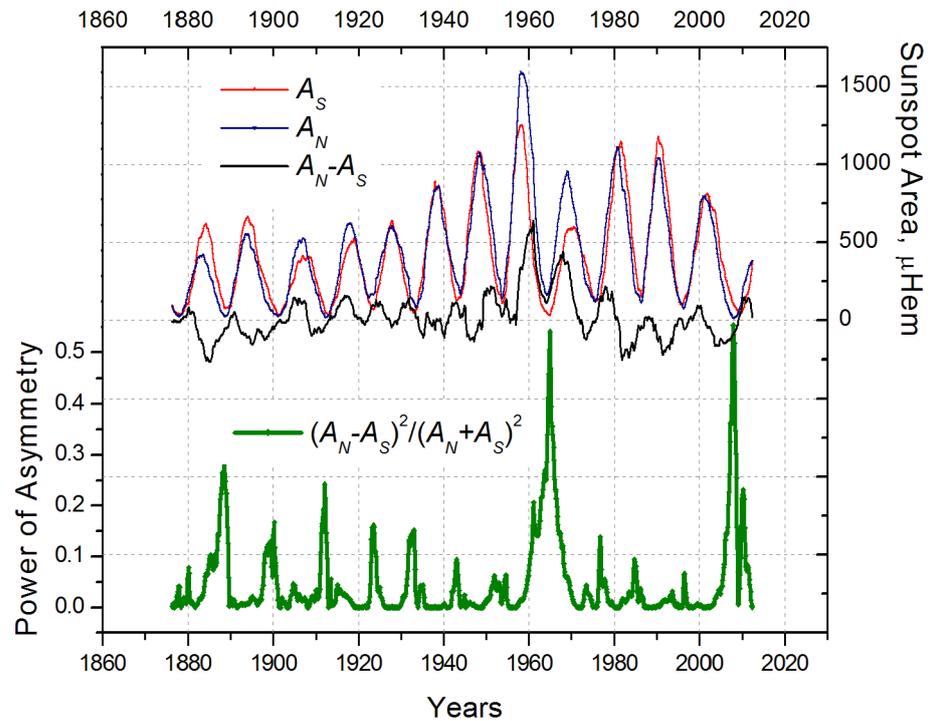

*Figure 1.* Greenwich sunspot areas for northern ($A_N$) and southern ($A_S$) hemispheres and their difference ($A_N - A_S$) (upper panel) and the power of the asymmetry (lower panel). Daily sunspot areas are smoothed in a 4-yr running window and digitized with a step of 50 days.

Specifically, daily sunspot areas for each hemisphere $A_S$, $A_N$ are smoothed with a cubic spline which minimizes the weighted sum $p \sum (y_i - S_i)^2 + (1-p) \int (D^2 S(t))^2 dt$ of errors between the data *y* and spline *S* and the norm of the second derivative of S. The smoothing parameter $p=1/(1+28^3/.06)$ corresponds to 28 day smoothing and so eliminates periods shorter than the period of Carrington rotation.

## 3. Causality Detection Method

The detection of causality from a time series is an intensively developing technique (Granger, 1988; Čenys, Lasiene and Pyragas, 1991; Sugihara *et al.*, 2012; Ma, Aihara and Chen, 2014; van Nes *et al.*, 2015) but rigorous proofs and respective conditions for the possibility of causality detection from empirical observables are still unknown. In the case of observed variables in a natural system the direction of causality shown with *e.g.* convergent cross-mapping (CCM) may fail. Some exceptions are found even for model systems with randomized initial conditions and averaged results (*e.g.* McCracken and Weigel, 2014). The CCM is mostly used during recent years but it is an essentially local method based on analysis of distances between a few nearest neighbors in pseudo-phase space (Takens's space, see *e.g.* Sauer *et al.*, 1991). A limited number of nearest neighbors is typical feature of CCM and similar (*L* and *M*) Takens's space based methods (Chicharro and Andrzejak, 2009; Andrzejak *et al.*, 2003). The earlier "global" conditional dispersion (CD) method (Čenys, Lasiene and Pyragas, 1991) is based on the same Takens's procedure. CD method calculates distances for all available scales of reconstructed pseudo-phase space contrary to aforementioned CCM and *L* and *M* methods which use smallest scales with a few nearest neighbors only. This CD method has an obvious advantage because more distances are calculated for medium and big scales. Bigger number of distances can be critical for the analysis of observed natural time series, because those are always limited in length and have noise at small scales. In this study, we develop this "global" method with the introduction of new parameters for the description of causality.

The CD method utilizes one observable (*e.g.* sunspot area) of an unknown dynamical system (*e.g.* the solar dynamo) which is reduced to a set of ordinary differential equations (ODEs). Unknown variables can be substituted by time shifts of the single observable variable and a pseudo-phase space can be reconstructed which preserves the topology of the actual phase-space according to Takens's theorem (*e.g.* Sauer *et al.*, 1991):

$$X^m(t) = m^{-1/2}\{x(i), x(i+\tau),...,x(i+\tau(m+1))\}, i=1,..,N . \qquad (2)$$

Here $\tau$ – time delay and $N$ – length of time series. Vector manifold $X^m(t)$ is topologically equivalent to an unknown phase space $M$ responsible for the dynamo in the northern hemisphere. In a similar way, another observable variable is used to reconstruct the second vector manifold:

$$Y^m(t) = m^{-1/2}\{y(i), y(i+\tau),..., y(i+\tau(m+1))\}, i = 1,..,N. \qquad (3)$$

The conditional dispersion is calculated using the Heaviside step function $\theta$. Conditional dispersion quantifies the dispersion of the points in the manifold $Y$ under condition that synchronous points on a manifold $X$ are located within an $\varepsilon$-ball:

$$\sigma_{XY}^m(\varepsilon) = \left(\frac{\sum_{i \neq j}\|Y^m(i) - Y^m(j)\|^2\, \theta(\varepsilon - \|X^m(i) - X^m(j)\|)}{\sum_{i \neq j}\theta(\varepsilon - \|X^m(i) - X^m(j)\|)}\right)^{1/2}. \qquad (4)$$

We note that $\sigma_{XY}^m(\varepsilon) \neq \sigma_{YX}^m(\varepsilon)$ and introduce measures quantifying the difference similar to the comparison of statistical distributions:

$$P = <2(\sigma_{XY}^M - \sigma_{YX}^M)^2 / \sigma_{XY}^M>_\varepsilon, \quad N = <2(\sigma_{XY}^M - \sigma_{YX}^M)^2 / \sigma_{YX}^M>_\varepsilon. \qquad (5)$$

Here $P$ and $N$ are Pearson's and Neuman's measure respectively (Deza and Deza, 2006). Another measure of synchronization

$$S = <1 - (\sigma_{XY}^M + \sigma_{YX}^M)/2>_\varepsilon, \qquad (6)$$

should discriminate cases of independence and complete synchronization.

## 4. Causality Detection for Model Systems

Primary testing of the CD method was performed by Čenys, Lasiene and Pyragas (1991) for coupled Hennon maps and these results were reproduced by Volobuev and Makarenko (2016). The coupled Hennon map is a classical chaotic system, but it does not reproduce any feature of solar activity so we try to use here the Van der Pol system as a test system. This is well-studied ordinary differential equation (ODE) describing a non-linear oscillator fist proposed as an electrical model of the heart (Van der Pol and Van der Mark, 1928).

Long-term Pulses of Dynamic Coupling between Hemispheres

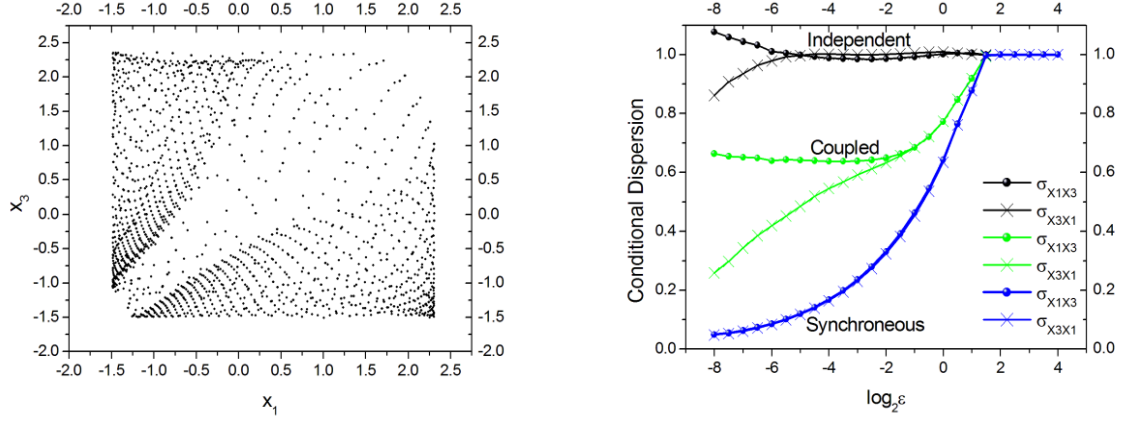

*Figure 2.* Solutions for coupled Van der Pol oscillators and the conditional dispersion. Left panel: scatter plot of variables tested in the case of weak coupling ($c_2$=0.02). Right panel: conditional dispersion given by Equation (4) with the scale $\varepsilon$ in the case of weak coupling (black), directional driving (green, $c_2$=0.15) and synchronization (blue, $c_2$=0.3)

It was shown that this type of equations can be derived via appropriate dimensionality reduction of the system of partial differential equations (PDEs) describing the oscillations of solar dynamo in an 11-yr cycle (Mininni, Gomez and Mindlin, 2001; Passos and Lopes, 2008). Although these both author groups used different technique, assumptions and simplifications, they get the same low-order dynamo model (LODM) described by the Van der Pol equation. Specifically, Passos and Lopes (2008) got the equation for the toroidal component of the magnetic field

$$\frac{d^2 B_\varphi}{dt^2} + \omega^2 B_\varphi + \mu\left(3\xi B_\varphi^2 - 1\right)\frac{dB_\varphi}{dt} - \lambda B_\varphi^3 = 0 \quad , \tag{7}$$

with oscillator parameters defined from solar parameters as

$$\omega^2 = \left(\frac{\mu}{2}\right)^2 - \frac{\alpha R \Omega}{\ell_0^2}, \quad \mu = 2\cdot\left(\frac{v_p}{l_0} + \frac{\eta}{l_0^2} - \frac{\eta}{R^2}\right), \xi = \frac{\gamma b_\varphi^2}{4\pi\rho\mu}, \lambda = \frac{\mu\gamma b_\varphi^2}{16\pi\rho}.$$

Here $\alpha$ is a constant describing the alpha-effect, $l_0$ is the characteristic length of interaction, $b_\varphi$ is the multiplier for $B_\varphi$ indicating its spatial profile, $R$ is the solar radius, $\Omega$ is the angular velocity, $v_p$ is the velocity of the meridional flow, $\eta$ is the magnetic diffusivity, $\gamma$ is a constant related to removal rate, and $\rho$ is the plasma density.

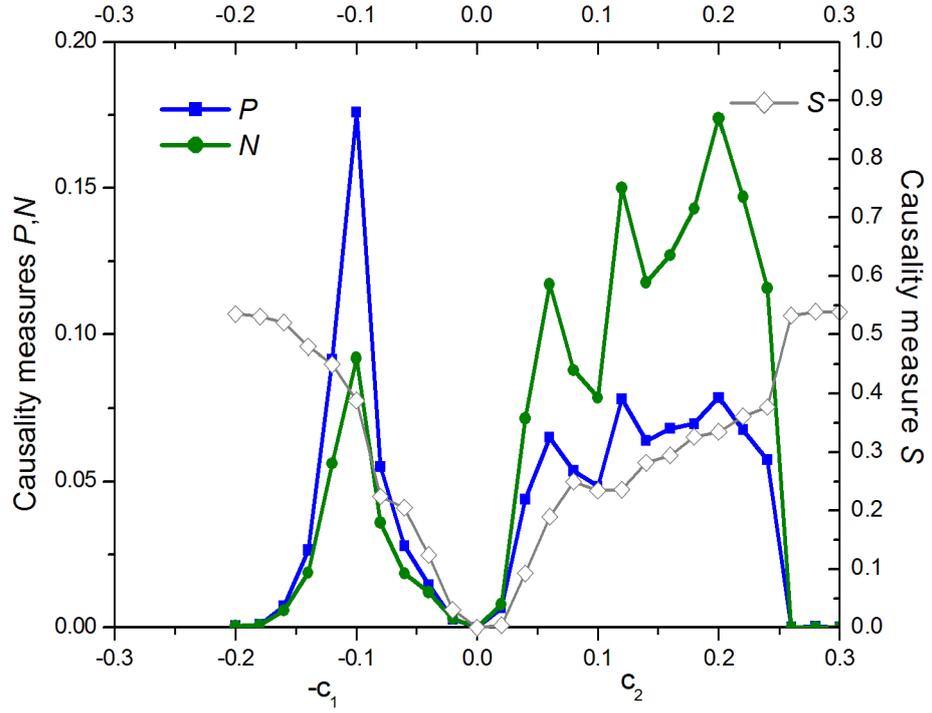

*Figure 3.* Causality measures for coupled Van der Pol oscillators. *P* (blue) and *N* (green) are Pearson and Neuman measures respectively, *S* (grey) is a measure of synchronization discriminating between independence and complete synchronization. $c_1$, $c_2$ are coupling parameters. Measure *N>P* means that X1→ X3 (X1 drives X3) and *vice versa*.

Most of the solar parameters for this zero-dimensional model are uncertain, and the model is oversimplified, so both Mininni, Gomez and Mindlin (2001) and Passos and Lopes (2008) fitted the parameters of the oscillator described by Equation (7) to each observed solar cycle separately and neglected cubic non-linearity ($\lambda = 0$).

Considering the hypothesis of a separate dynamo for each hemisphere, we use this general idea to construct the coupled system for the two hemispheres with the main purpose to illustrate the applicability of causality detection method to systems of this kind rather than to build a dynamo model.

Coupled Van der Pol equations can produce a variety of solutions including chaotic and periodic ones (dos Santos, Lopes and Viana, 2004). The weakly coupled equations produce a chaotic solution (Figure 2). The general non-dimensional Lienard form for coupled Van der Pol systems is as follows:

$$\frac{dx_1}{dt} = x_2, \frac{dx_2}{dt} = k(x_1 - w_1)(x_1 - w_2)x_2 - b_1 x_1 + c_1(x_3 - x_1), \tag{8}$$

Long-term Pulses of Dynamic Coupling between Hemispheres

$$\frac{dx_3}{dt} = x_4, \frac{dx_4}{dt} = k(x_3 - w_1)(x_3 - w_2)x_4 - b_2 x_3 + c_2(x_1 - x_3). \quad (9)$$

Here the constants $b_1 = 1$, $b_2 = 1.01$, $k = 1.45$, $w_1 = -0.2$, $w_2 = 1.9$ are taken from dos Santos, Lopes and Viana (2004), to provide variability of solution for unforced oscillators changing from a chaotic to a synchronized state with an increase of coupling parameter. Interpreting the results, we can consider $x_1$ as toroidal field in the nothern hemisphere, $B_{\varphi N}$, but with the caution that Equation (8) has the additional term $x_1 \frac{dx_1}{dt}$ whereas Equation (7) has no such term.

With this caution the coupling here can be interpreted as the magnetic diffusion of toroidal field between hemispheres through the equator. It seems that this effect is observed on the Sun (McIntosh *et al.*, 2015). Initial conditions are set equal for both systems ($x_1(0)=2$, $x_2(0)=-1$, $x_3(0)=2$, $x_4(0)=-1$). Avoiding dependence on initial conditions we analyze stabilized solutions at 200<*t*<500, digitized to 1500 points, which gives approximately 50 points per period. This model data coverage is close to the actual time window with 1460 daily points (4 years) which is used further to analyze the sunspot area time series. One of the coupling parameters $c_1$, $c_2$ is varied while the other one is set equal to zero, simulating a unidirectional coupling. Solutions for $x_1(t)$ and $x_3(t)$ are found via integration of Equations (8)-(9) with a Runge-Kutta 4[th] order scheme. These solutions are used to reconstruct the respective Takens's space, Equations (2)-(3), for the "shadow" systems *X1* and *X3*. Measures *P* and *N*, Equation (5), allow estimation of the directionality of the coupling for each value of the coupling parameter (Figure 3). The case *N>P* shows detectable directionality of coupling *X1*→*X3* and *vice versa*. If $c_2$ is too big ($c_2$>0.25) the coupling is stronger but the directionality of coupling becomes undetectable and the system goes into its synchronized state. Measure *S*, Equation (6), shows the strength but not the directionality of coupling. The MATLAB codes for the algorithm with a described example are available at https://www.mathworks.com/matlabcentral/fileexchange/61076-conditional-dispersion-to-detect-causality-between-chaotic-time-series.

## 5. Causality Detection for Solar Hemispheres

Using the approach of Section 4 which indicates the coupling between the model systems, we apply the CD method to detect the time-dependent causality between the magnetic activities in the solar hemispheres. Preprocessed (Section 2) daily sunspot areas are used as observables to reconstruct the shadow manifolds (Equations (2)-(3)) for each solar hemisphere using the time delay $\tau = 50$ days. Conditional dispersion (Equation (4)) and measures (Equations (5)-(6)) are calculated in a 4-yr moving window with a 50 day step. It appears that the hemispheres are synchronous most of the time, so that the synchronization index $S$ (Equation (6), gray curve in Figure 4) is close to its highest value (about 0.55) observed in the model Van der Pol system, although it has a long-term trend which repeats the trend of the total sunspot area. An important feature is the appearance of some peaks with decreasing $S$, which are probably of two types having different physics. The first type, *e.g.* the peak around the year 1900, has pronounced peak of lost synchronization near the minimum of solar activity but has no driving, so that the difference between Pearson and Neumann measures is very small (Figure 4, blue and green curves). The hemispheres suddenly lose the synchronization but no driving appears at that moment, at least as indicated by sunspot areas. Another type is *e.g.* the peak at the recent minimum of activity between Cycles 23 and 24. The synchronization $S$ decays simultaneously with the growth of both the Pearson and Neumann measures, and the growth of the difference between them, indicating a directional coupling between the hemispheres.

Long-term Pulses of Dynamic Coupling between Hemispheres

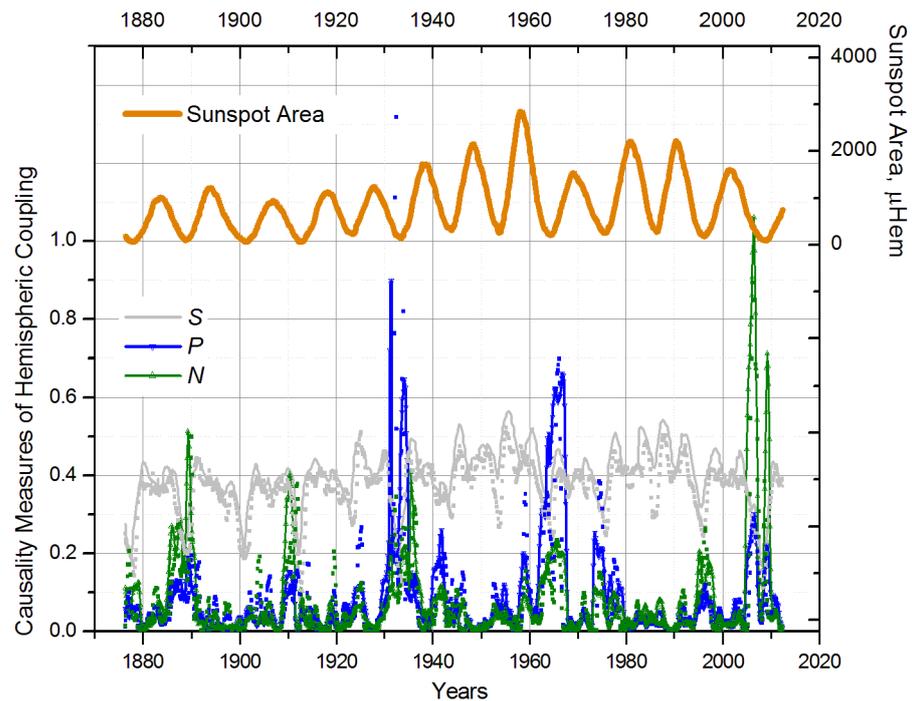

*Figure* 4. Different measures of hemispheric coupling as derived from the conditional dispersion (lower panel) and total sunspot area (upper panel) calculated in with a 4yr moving window. The difference between the Pearson (*P*) and Neumann (*N*) measures (blue and green curves) shows the presence of directional coupling between the hemispheres. The synchronization measure (*S*) shows the presence of synchronization similar to Figure 2. The scattered points show the same measures calculated with a 2yr moving window.

The peaks of the second type are relatively regular phenomenon during the period of the Greenwich observations (Figure 2). They do not repeat at each 11-year cycle nor even at each 22-year cycle. It appears that this is a more pronounced manifestation of some longer-term cycles, than we have seen in the statistics of solar indices before. These peaks cannot be attributed to statistical noise. Probably they confirm the presence of strong directional coupling between the hemispheres during some minima of solar activity. The direction of coupling can be seen from Figure 4 by comparison of the Pearson and Neumann measures (blue and green curves). More directly we quantify the direction of coupling as the difference between these measures, $N - P$, as shown in Figure 5. Directionality peaks can be attributed to big enough peaks in the power of asymmetry. After these peaks hemispheres return to a synchronous dynamics there the direction of coupling is close to zero (Figure 5) so that the direction is difficult to determine. Probably, actual sign of the direction after certain peak is preserved until the peak of opposite direction will appear as indicated in Figure 1, blue and red horizontal bars. An exception is the fast transition of the determined direction of coupling with a changing of sign during 1934. In that case the power of asymmetry is small but direction of coupling changes without transition through the synchronous regime.

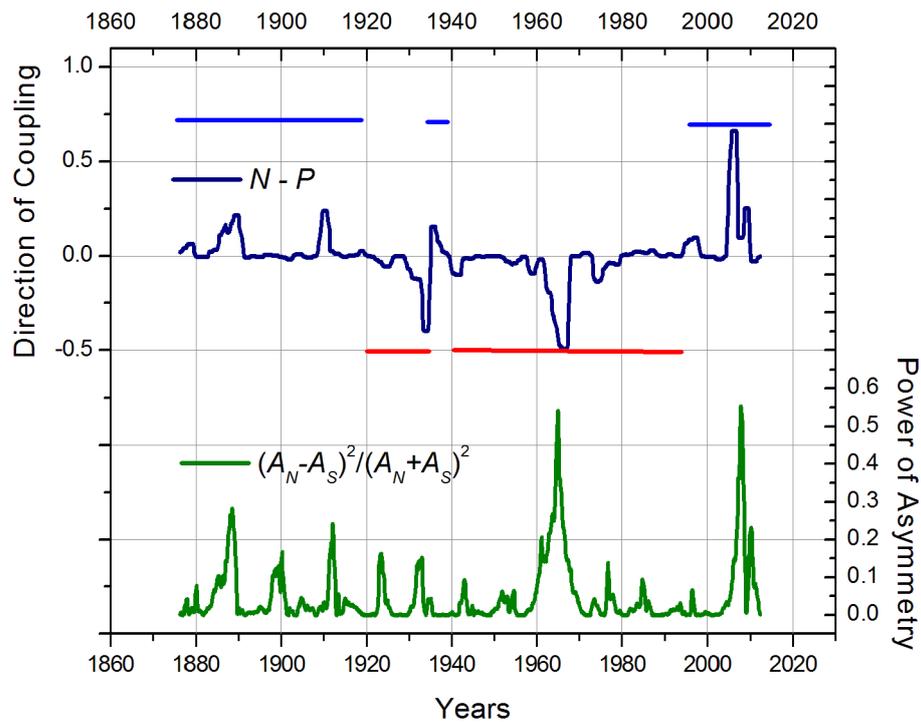

*Figure* 5. Direction of coupling between the solar hemispheres (upper panel) and the power of asymmetry (lower panel) calculated with a 4yr moving window. Blue and red horizontal bars indicate periods of dominant directionality of coupling. Spikes are removed with a 2yr median filter.

We observed one relatively long period of unidirectional coupling of more than 40 years from 1940 until at least 1980 year (Figure 5). Although pulses of significant directionality during this period are relatively short, they all have the same negative sign and probably return to a synchronized state with the same direction of causality. This phenomenon reveals a memory for at least two Hale pairs of cycles.

## 6. Discussion and Conclusions

We apply here the causality detection method known from nonlinear chaotic dynamics to sunspot area time-series in each hemisphere. Despite a relatively

Long-term Pulses of Dynamic Coupling between Hemispheres

short 4yr moving window used for calculation of the causality measures, we find that the direction of causality reveals no 11-year cycle but local pulses repeat at the rate of the Hale cycle or longer.

This is non-trivial result because a 4yr moving average does not suppress the 11-year cycle in sunspot area nor asymmetry indices. The pulses repeating after three and four 11-year cycles provide direct observational proof for the existence of long-term memory in the solar dynamo. This long-term memory is not directly seen in observed activity indices. If we observe the correlation of the $n^{th}$ Cycle polar fields with the next $(n+1)^{th}$ Cycle amplitude, it will correlate with the $(n+1)^{th}$ Cycle but not with the $(n+2)^{nd}$ or $(n+3)^{rd}$ cycle. This is true for other observable parameters (see *e.g.*, Jiang *et al*. 2007; Karak and Choudhuri, 2011; Hazra *et al.* 2015). We can statistically prove the possibility of prediction for the one decadal number of the envelope but not for two numbers (Volobuev and Makarenko, 2008). Thus besides the highly irregular shape of the Gleissberg cycles the appearance of repeating causality peaks is probably the first evidence for a long-term memory in the Sun.

**Acknowledgements** We thank anonymous referee for useful discussions and suggestions which greatly improved the manuscript. The work was supported by Russian Foundation for Basic Research, Project No. 15-01-09156
**Disclosure of Potential Conflicts of Interest** The authors declare that they have no conflicts of interest.

# References

Andrzejak, R. G., Kraskov, A., Stögbauer, H., Mormann, F., Kreuz, T.: 2003, Bivariate surrogate techniques: necessity, strengths, and caveats. *Phys. Rev. E*, 68(6), 066202 DOI: https://doi.org/10.1103/PhysRevE.68.066202

Bushby, P. J., Tobias, S. M.: 2007, On predicting the solar cycle using mean-field models. *Astrophys. J.*, 661(2), 1289.

Carbonell, M., Oliver, R., Ballester, J. L.: 1993, On the asymmetry of solar activity. *Astron. Astrophys.*, 274, 497. http://adsabs.harvard.edu/abs/1993A&A...274..497C

Carbonell, M., Terradas, J., Oliver, R., Ballester, J. L.: 2007, The statistical significance of the North-South asymmetry of solar activity revisited. *Astron. Astrophys.*, 476(2), 951-957. DOI: 10.1051/0004-6361:20078004

Čenys, A., Lasiene, G., Pyragas, K.:1991, Estimation of interrelation between chaotic observables. *Physica D: Nonlinear Phenomena*, 52(2), 332-337. DOI: https://doi.org/10.1016/0167-2789(91)90130-2

Chatterjee, P., Choudhuri, A. R.: 2006, On magnetic coupling between the two hemispheres in solar dynamo models. *Solar Phys.*, 239(1-2), 29-39. DOI: 10.1007/s11207-006-0201-6

Chatterjee, P., Nandy, D. and Choudhuri, A. R.: 2004, Full-sphere simulations of a circulation-dominated solar dynamo: Exploring the parity issue. *Astron. Astrophys.*, 427(3), 1019-1030. https://doi.org/10.1051/0004-6361:20041199

Chicharro, D., Andrzejak, R. G.: 2009, Reliable detection of directional couplings using rank statistics. *Phys. Rev. E*, 80(2), 026217 DOI:https://doi.org/10.1103/PhysRevE.80.026217

Deza, M. M., Deza, E.: 2006, *Dictionary of distances.* Elsevier, Amsterdam.

dos Santos, A. M., Lopes, S. R., Viana, R. R. L.: 2004, Rhythm synchronization and chaotic modulation of coupled Van der Pol oscillators in a model for the heartbeat. *Physica A: Statistical Mechanics and its Applications*, 338(3), 335-355. http://dx.doi.org/10.1016/j.physa.2004.02.058

Granger, C. W.: 1988, Some recent development in a concept of causality. *Journal of Econometrics*, 39(1), 199-211. https://doi.org/10.1016/0304-4076(88)90045-0

Hazra, G., Karak, B. B., Banerjee, D., and Choudhuri, A. R.: 2015, Correlation between decay rate and amplitude of solar cycles as revealed from observations and dynamo theory. *Solar Phys.*, 290(6), 1851-1870. DOI: 10.1007/s11207-015-0718-8

Hotta, H., Yokoyama, T.: 2010, Solar parity issue with flux-transport dynamo. *Astrophys. J. Lett.,* 714(2), L308. DOI: 10.1088/2041-8205/714/2/L308

Javaraiah, J.: 2015, Long-term variations in the north–south asymmetry of solar activity and solar cycle prediction, III: Prediction for the amplitude of solar cycle 25. *New Astron.*, 34, 54-64. http://dx.doi.org/10.1016/j.newast.2014.04.001

Jiang, J., Chatterjee, P. and Choudhuri, A. R.: 2007, Solar activity forecast with a dynamo model. Monthly Notices of the Royal Astronomical Society, 381(4), 1527-1542. DOI: https://doi.org/10.1111/j.1365-2966.2007.12267.x

Olemskoy, S. V., Kitchatinov, L. L.: 2013, Grand minima and North-South asymmetry of solar activity. *Astrophys. J.*, 777(1), 71. doi:10.1088/0004-637X/777/1/71

Karak, B. B., Choudhuri, A. R.: 2011, The Waldmeier effect and the flux transport solar dynamo. Monthly Notices of the Royal Astronomical Society, 410(3), 1503-1512. DOI: https://doi.org/10.1111/j.1365-2966.2010.17531.x

Ma, H., Aihara, K., Chen, L.: 2014, Detecting causality from nonlinear dynamics with short-term time series. *Scientific reports*, 4, 7464. DOI: 10.1038/srep07464

Mandal, S. , Hegde, M., Samanta, T., Hazra, G., Banerjee, D., Ravindra B.: 2017,

Kodaikanal digitized white-light data archive (1921-2011). Analysis of various solar cycle features. *Astron. Astrophys.* https://doi.org/10.1051/0004-6361/201628651

McCracken, J. M., Weigel, R. S.: 2014, Convergent cross-mapping and pairwise asymmetric inference. *Physical Review E*, 90(6), 062903.DOI:https://doi.org/10.1103/PhysRevE.90.062903

McIntosh, S. W., Leamon, R. J., Krista, L. D., Title, A. M., Hudson, H. S., Riley, P., Harder, J. W., Kopp G., Snow, M., Woods, T. N., Kasper, J. C., Stevens, M. L., Ulrich, R. K.: 2015, The solar magnetic activity band interaction and instabilities that shape quasi-periodic variability. *Nature Communications*, 6, 6491. DOI: 10.1038/ncomms7491

Long-term Pulses of Dynamic Coupling between Hemispheres

Mininni, P. D., Gomez, D. O., Mindlin, G. B.: 2001, Simple model of a stochastically excited solar dynamo. *Solar Phys.*, 201(2), 203-223; DOI: 10.1023/A:1017515709106

Passos, D., Lopes, I.: 2008, A low-order solar dynamo model: inferred meridional circulation variations since 1750. *Astrophys. J.*, 686(2), 1420. http://iopscience.iop.org/article/10.1086/591511/meta

Sauer, T., Yorke, J. A., Casdagli, M.: 1991, Embedology. *J. of Stat. Phys.*, 65(3-4), 579-616. doi:10.1007/BF01053745

Shetye, J., Tripathi, D., Dikpati, M. (2015). Observations and modeling of north-south asymmetries using a flux transport dynamo. *Astrophys. J.*, 799(2), 220. DOI: 10.1088/0004-637X/799/2/220

Sokoloff, D., Nesme-Ribes, E.: 1994, The Maunder minimum: A mixed-parity dynamo mode? *Astron. Astrophys.*, 288, 293-298. http://adsabs.harvard.edu/abs/1994A&A...288..293S

Sugihara, G., May, R., Ye, H., Hsieh, C. H., Deyle, E., Fogarty, M., & Munch, S.: 2012, Detecting causality in complex ecosystems. *Science*, 338 (6106), 496-500. DOI: 10.1126/science.1227079

Temmer, M., Rybák, J., Bendík, P., Veronig, A., Vogler, F., Otruba, W., ... , Hanslmeier, A. (2006). Hemispheric sunspot numbers Rn and Rs from 1945-2004: catalogue and N-S asymmetry analysis for solar cycles 18-23. *Astron. Astrophys,* 447(2), 735-743. DOI: 10.1051/0004-6361:20054060

Van der Pol, B., Van der Mark, J.: 1928, LXXII. The heartbeat considered as a relaxation oscillation, and an electrical model of the heart. The London, Edinburgh, and Dublin Philosophical Magazine and Journal of Science, 6(38), 763-775.

van Nes, E. H., Scheffer, M., Brovkin, V., Lenton, T. M., Ye, H., Deyle, E., Sugihara, G.: 2015. Causal feedbacks in climate change. *Nature Climate Change*. DOI: 10.1038/NCLIMATE2568

Volobuev, D.: 2006, "TOY" Dynamo to Describe the Long-Term Solar Activity Cycles. *Solar Phys.*, 238(2), 421-430. DOI: 10.1007/s11207-006-0154-x

Volobuev, D. M., Makarenko, N. G.: 2008, Forecast of the decadal average sunspot number. *Solar Phys.*, 249(1), 121-133. DOI: 10.1007/s11207-008-9167-y

Volobuev, D.M., Makarenko, N.G.: 2016, The Dynamic Relation between Activities in the Northern and Southern Solar Hemispheres. *Geomagnetism and Aeronomy*, 56 (7), 880-885 DOI: 10.1134/S0016793216070173